# The Relative Performance of Filled and Feedhorn-Coupled Focal Plane Architectures

Matthew J. Griffin[1], James J. Bock[2], and Walter K. Gear[1]

[1] Department of Physics and Astronomy, University of Wales, Cardiff, Queens Buildings, PO Box 913, 5 The Parade, Cardiff CF24 3YB, UK

[2] Jet Propulsion Laboratory, California Institute of Technology, Pasadena CA91109, USA

**Abstract**

Modern far infrared and submillimeter instruments require large format arrays. We consider the relative performance of filled array (bare pixel) and feedhorn-coupled architectures for bolometer focal planes. Based on typical array parameters, we quantify the relative observing speeds and comment on the merits of the different architectures. Filled arrays can provide higher mapping speed (by a factor of up to 3.5) and simpler observing modes at the expense of reduced sensitivity for pointed observations, increased detector numbers, and greater vulnerability to stray light and electromagnetic interference. Taking advantage of the filled array architecture requires strongly background limited detectors. At millimeter wavelengths, filled arrays must be surrounded by a sufficiently cold enclosure to minimize the background power from the instrument itself.

## 1. Introduction

Imaging bolometer instruments in the submillimeter spectral region ($l > 200$ μm) are needed for a variety of important astrophysical studies including star formation, galaxy formation and evolution, active galactic nuclei, evolved stars, comets and asteroids. To date, most bolometer camera instruments have used composite semiconductor bolometers mounted in individual integrating cavities and coupled to the telescope by means of single-mode or few-moded feedhorns. New bolometer instruments are now being built or planned to provide improved sensitivity and field of view. An alternative focal plane array architecture involves dispensing with feedhorns and filling the focal plane with an array of bare bolometers. For a bolometric detector, the overall sensitivity can be characterized by the combination of two uncorrelated contributions to the Noise Equivalent Power (NEP): that due to the detector itself and that due to the unavoidable statistical fluctuations in the background power incident on the detector from the thermal radiation of the instrument, the telescope, the atmosphere, or any combination of these. Ideally, the NEP due to the background power dominates, and the sensitivity is said to be background limited..

In this paper we consider the advantages and disadvantages of these alternative array architectures and calculate their relative observing speed for pointed and mapping observations, based on reasonable assumptions and typical array parameters. We calculate the speed improvements achievable in principle (for the completely background-limited case), and also consider the effect of significant detector NEP on observing speed. The feedhorn and filled array architectures, and the assumptions made for the purposes of inter-comparison, are described in Section 2. The main performance parameters of the different options are presented in Section 3.

In Section 4 we derive expressions for the relative observing speed of the different options, and the results of calculations are presented and discussed in Section 5.

The definitions of the symbols used in this paper are given below.

| | |
|---|---|
| $A_T$ | Telescope area |
| $A\boldsymbol{W}$ | Throughput (area-solid angle product) with which the detectors receive incident radiation |
| $B_{\boldsymbol{n}}(T)$ | Planck function |
| $B_{ext}$ | Overall surface brightness viewed by the detectors coming from outside the instrument |
| $B_{int}$ | Overall surface brightness viewed by the detectors coming from the cold box surrounding the detectors |
| $D$ | Telescope primary aperture diameter |
| $F$ | Focal ratio of the instrument final optics feeding the detectors |
| $f$ | Factor by which point source observing speed can be improved by co-adding the central and eight neighboring pixels for a $0.5F\boldsymbol{l}$ filled array |
| $N$ | Number of detectors in an array of a given total area |
| $NEP_{det}$ | Detector $NEP$ |
| $NEP_{ph}$ | Photon noise limited $NEP$ referred to the background radiant power absorbed by the detector |
| $P_{A, C, S, T}$ | Power absorbed by a detector from the atmosphere, instrument cold box, astronomical sky, and telescope, respectively |
| $P_{tot}$ | Total radiant power absorbed by a detector |
| $S_\nu$ | Point source flux density at the telescope aperture |
| $T_{A, C, S, T}$ | Atmospheric temperature, detector cold box temperature, astronomical sky brightness temperature, and telescope temperature, respectively |
| $g$ | Ratio of detector *NEP* to photon noise *NEP* (i.e., $NEP_{det}/NEP_{ph}$) |
| $s$ | Signal-to-noise ratio for a given detector or map point |
| $\Delta q$ | Angular offset of the telescope pointing from a point source |
| $e_A$ | Atmospheric emissivity |
| $e_T$ | Telescope emissivity |
| $\lambda, \nu$ | Central wavelength, frequency of band-pass filter |
| $\Delta\lambda, \Delta\nu$ | Filter bandwidth (the filter is assumed to have a top-hat transmission profile) |
| $\eta_A$ | Aperture efficiency: this is the fraction of the total power from a point source diffraction pattern that is coupled to a detector centered on the source |
| $\eta_d$ | Detector responsive quantum efficiency |
| $\eta_o$ | Instrument overall transmission efficiency |
| $\eta_s$ | Spillover efficiency: this is the fraction of the detector throughput which illuminates the telescope (a fraction $1 - \eta_s$ is assumed to terminate on the inside of the detector cold box) |
| $\theta_{FWHM}$ | Full-width-half-maximum beam width on the sky |
| $t$ | Integration time per detector |

## 2. Array architectures

### 2.1 Feedhorn arrays

Most submillimeter bolometer cameras to date have used feedhorn arrays with NTD germanium composite bolometers [e.g., Refs. 1, 2, 3, 4]. Feedhorn arrays will also be used for the Herschel-SPIRE [5] and Planck-HFI [6] satellite instruments.

Circular-aperture feedhorns are usually close-packed in the focal plane to fit as many detectors as possible into the available area. Corrugated horns provide a near Gaussian antenna pattern.[7] Smooth-walled conical horns have higher side-lobes due to the sharp transition in the fields at the outer edge of the horn aperture, but are simpler and cheaper to manufacture in large numbers. The feedhorn restricts the detector beam solid angle, giving a tapered illumination of the telescope primary mirror, with an edge taper of typically 8-12 dB for $2F\lambda$ horns. Down to a level of 20 dB or more the main beam profile on the sky is a good approximation to a Gaussian[8], whose width depends on the edge taper. Observing modes with feedhorn arrays must take into account the fact that although the horns are close-packed in the focal plane, their beams on the sky do not fully sample the image unless the horn diameter is less than the Nyquist sampling interval of $0.5F\lambda$, since the telescope aperture diameter in wavelengths band-limits accepted angular frequencies to those less than $D/\lambda \equiv 1/(F\lambda)$ in the focal plane.

Several separate telescope pointings are therefore needed to create a fully-sampled image. For $2F\lambda$ horns, at least 16 pointings are required, as illustrated in Figure 1. For horns of $1F\lambda$ aperture, 4 pointings are required. Full sampling is achieved by moving ("jiggling") the array pointing.[1] Alternatively, a fully sampled image of a large area can be obtained by scanning the array across the sky at the appropriate angle to provide the necessary overlap between beams.

The main advantages of $2F\lambda$ feedhorn arrays are:

(i) a $2F\lambda$ feedhorn provides maximum efficiency for detection of a point source with known position (because most of the power from the source is concentrated on a single detector);

(ii) horn properties are well-understood, allowing good control of the beam and reliable design;

(iii) the bolometer angular response is restricted to the telescope, giving good stray light rejection;

(iv) the susceptibility to electromagnetic interference can be controlled - the horn plus integrating cavity act as a Faraday enclosure;

(v) the number of detectors needed to fill a given array field of view is minimized.

The main disadvantages are:

(i) in order to achieve full spatial sampling of the sky, even for a region smaller than the array field of view, jiggling or scanning are needed, which complicates the observing modes;

(ii) the efficiency for mapping is considerably less than the ideal value.

The second disadvantage is the more fundamental and important one. The interferometric nature of any antenna means that an inevitable price is paid for its directivity in that some of the power incident on the horn array is actually reflected back out. Another way of regarding this is that the feedhorn couples only to the fundamental mode and does not respond to the significant incident power contained in the higher order modes of the incident beam.

**2.2 Filled arrays**

Filled detector arrays dispense with feedhorn or antenna coupling of the detectors to the incoming beam in favor of an array of rectangular (usually square) absorbing pixels. To achieve instantaneous full sampling of the image, the pixel center-to-center spacing must be $0.5F\lambda$ or less. The broad angular response of each individual pixel (~ $\pi$ sr), requires the use of a cold stop in the optical system, and gives a flatter illumination profile for the telescope. This is advantageous if the application requires maximum point source sensitivity, but not if high side-lobe suppression is needed (this configuration might not be optimal for cosmic background observations, for instance). For composite bolometric detectors, maximum pixel absorption efficiency requires that a reflecting back-short plate be positioned at $\lambda/4$ behind the array. The need to restrict the physical size of the array while maintaining a large pixel number limits the pixel size, which can be comparable to the wavelength. For an isolated detector this would result in a low absorption efficiency due to diffraction effects. However, for a filled array, capacitive coupling between the pixels should result in a high overall efficiency for the array even when the pixels are smaller than the wavelength, with the array appearing to the incident beam as a continuous resistive sheet. Thus, the radiation from the telescope arriving at the focal plane can be absorbed with very high efficiency, and this means that there is a potential increase in sensitivity over the feedhorn antenna coupled detector. Electromagnetic modelling of filled array architectures has indicated that this high absorption efficiency can be achieved.[9]

The main advantages of filled arrays (relative to $2F\lambda$ feedhorn arrays) are:

(i) they provide higher efficiency for mapping observations (as quantified in Section 5 below);

(ii) full sampling of the instantaneous field of view of the array can be achieved by using pixels of $0.5F\lambda$ or smaller, obviating the need for jiggling;

(iii) for a $0.5F\lambda$ array, the beam profile on the sky can be slightly narrower for a given telescope size due to the stronger illumination of the outer parts of the telescope.

Disadvantages are:

(i) the background power per pixel is lower than for the larger feedhorn-coupled detectors, typically by a factor of 4 - 5 (for the case considered below in section 3.4, the factor is 4.8), giving a photon noise *NEP* which is lower by a factor of 2 or more, and thus more difficult to achieve;

(ii) the detectors are much more vulnerable to stray light due to the very broad pixel angular response - by a factor of $\pi F^2/4$, assuming a pixel beam solid angle of $\pi$ sr;

(iii) the vulnerability to electromagnetic interference is also greater due to the "naked" array architecture;

(iv) more detectors are needed to fill a given field size: it can be shown that, in the large field limit, an array of $0.5F\lambda$ square pixels has $16\cos(30°) = 13.9$ times more detectors than an array of close-packed $2F\lambda$ circular feedhorns, and $4\cos(30°) = 3.46$ times more detectors than a $1F\lambda$ feedhorn array.

To date, the only operating instrument which has adopted this approach is SHARC [10], in use on the Caltech Submillimeter Observatory (CSO). Filled bolometer arrays will be implemented in SHARC-2 [11], Sofia-HAWC [12], Herschel-PACS [13], and SCUBA-2 [14].

### 2.3 Assumptions

Filled array pixels larger than $0.5F\lambda$, or feedhorns smaller than $2F\lambda$, may also be used as a compromise between the extreme cases discussed above. In this work we consider the following four configurations: filled array with $0.5F\lambda$ pixels; filled array with $1F\lambda$ pixels; feedhorn array with $1F\lambda$ pixels; feedhorn array with $2F\lambda$ pixels. For ease of comparison, the performance characteristics of the other three options are normalized with respect to the $2F\lambda$ feedhorn case. In estimating the theoretical relative performance of the different array architectures, we make the following assumptions:

1. The point spread function (PSF) is purely diffraction limited. This will not be so in a real instrument but can be closely approximated in practice at submillimeter wavelengths [e.g., Ref. 15]. The assumption is effectively that the influence of imperfect PSF on sensitivity will be the same for both feedhorn and filled array options.

2. Beam profiles on the sky can be approximated by Gaussians in each case.

3. Emission from four sources can contribute to the total power absorbed by the detectors: the telescope, the Earth's atmosphere (if appropriate), the instrument cold box, and the astronomical sky. While the last of these can be significant for cosmic background observations, in the case of point source observations, we take the contribution of the point source itself to be negligible.

4. The photon noise level is closely approximated by the Poissonian contribution (photon shot noise) - i.e., the Bose-Einstein term in the photon noise limited NEP [e.g., Ref. 16] can be neglected. This simplifies the calculations and is reasonably well approximated in real submillimeter instruments.

5. Detector center-center spacings are $0.5F\lambda$ or $1F\lambda$ for the filled array case and $1F\lambda$ or $2F\lambda$ for the feedhorn case. To account for a finite inter-pixel gap or horn wall thickness, the active pixel sizes are assumed to be 5% less than the center-center distance: $0.475F\lambda$, $0.950F\lambda$, and $1.90F\lambda$.

6. Filled array pixels are square and have a broad angular response which we take to be characterized by a solid angle of $p$ sr. A small portion of this solid angle, determined by the focal ratio of the final optics, $F$, views the telescope through an aperture in the cold box.

7  Feedhorns are single-moded (throughput $A\Omega = \lambda^2$) .

8  The spectral passband is determined by a top-hat filter centered on wavelength $\lambda$ with a bandwidth $\Delta\lambda$.

9  The optical efficiency of the instrument, the absorption efficiency of the detectors, and the inherent detector NEP are the same for all options.

10  There is no stray light or out-of-band radiation.

11  Observing overheads are negligible or equivalent for the different options.

Throughout, subscript 0 is used to refer to the case of $2F\lambda$ feedhorns, against which the other array options are compared.

## 3. Aperture efficiency, throughput, and beam profiles

### 3.1 Aperture efficiency for filled array pixels

For an individual square pixel in a filled array, $\eta_A$ is determined by the fraction of the power in the PSF that is contained within the pixel area, which may be calculated by integrating under the intercepted part of the PSF. This is plotted against pixel side in Figure 2 for a diffraction limited system (where the PSF is the Airy diffraction pattern). For 5% inter-pixel gap, $\eta_A$ is 0.162 for the $0.5F\lambda$ case, and 0.495 for the $1F\lambda$ case.

### 3.2 Aperture efficiency for feedhorns

Figure 2 also shows $\eta_A$ vs. horn diameter (in units of $F\lambda$) for a smooth-walled conical horn.[17] (This depends to some extent on the horn length - this curve is for an "aperture-limited" horn: i.e., the horn is sufficiently long that the beam profile is a good approximation to that of an infinitely long horn). The maximum feedhorn aperture efficiency is about 0.75 for an aperture close to $2Fl$. In practice, a peak value of $h_{A0} = 0.7$ is more realistic for the $2Fl$ feedhorn, taking into account finite horn length and wall thickness.[18] For diameters above $2Fl$, the efficiency begins to decrease because, as the horn aperture increases, its angular response gets narrower so that the outer parts of the telescope are illuminated less efficiently thereby reducing the effective collecting area. This is in contrast to the filled array case - here the aperture efficiency continues to increase as the pixel (a Lambertian absorber) gets bigger and absorbs more and more energy in proportion to the intercepted portion of the PSF.

For $1Fl$ horns, we adopt a value of $h_A = 0.35$, also somewhat lower than the theoretical maximum. Although reducing the horn diameter lowers the efficiency of each horn, more horns can be accommodated in a given focal plane area. Smaller horns have a broader angular response; the pupil stop fills a smaller proportion of the throughput with a correspondingly lower angular taper on the pupil, but a larger fraction of the throughput is then terminated on the cold box.

### 3.3 Throughput and background power per pixel for filled array pixels

For square pixels with an angular response of total solid angle $p$ sr, the throughput is:

for 0.5$Fl$ pixels with 5% gap $\qquad A W = [(0.5)(0.95)Fl]^2 p = 0.226 F^2 l^2 p$ ; (1)

for 1$Fl$ pixels with 5% gap $\qquad A W = [(1.0)(0.95)Fl]^2 p = 0.903 F^2 l^2 p$ . (2)

For low values of $F$, the throughput per pixel for a filled array can be comparable to or even less than $l^2$, the single-moded feedhorn value. However, the distribution (area - solid angle) of the pixel throughput is very different in the two cases: the filled array has a broad angular response regardless of the value of $F$. The filled array pixel can be regarded as a collector of photons in the focal plane, with a rate that simply decreases in proportion to its area. Most of the filled array throughput views the cold box, and the pixels view the telescope and sky with a top-hat beam of solid angle $p/(4F^2)$, corresponding to a fraction of the total throughput of $h_s = 1/(4F^2)$. The throughput for the external background radiation is thus $0.177l^2$ for $0.5Fl$ pixels and $0.709l^2$ for $1Fl$ pixels.

The power per pixel scales with $AW$, so the $1Fl$ pixel receives four times more background power than the $0.5Fl$ pixel. The photon noise limited $NEP$ is proportional to the square root of the background power and is consequently higher for the $1Fl$ pixel by a factor of 2.

### 3.4 Throughput and background power per pixel for feedhorns

In the case of a feedhorn-coupled detector, the illumination of the telescope is approximately Gaussian, with an edge taper on the pupil that depends on the feedhorn antenna pattern and the details of the optical system. For single-moded feedhorns, the total horn throughput is $AW = l^2$. A fraction $h_s$ views the telescope and sky and a fraction $(1 - h_s)$ is terminated on the cold box. Figure 3 shows the spillover efficiency, $h_s$, as a function of the edge taper for a Gaussian illumination of the pupil. Typical values for the edge taper would be in the range 7 - 10 dB for $2Fl$ horns and 1.5 - 3 dB for $1Fl$ horns. In this work we assume an edge taper for $2Fl$ of 8 dB, corresponding to $h_{s0} = 0.83$. To first order, the main beam of a $1Fl$ feedhorn will be twice as wide as that of the $2Fl$, which corresponds to a 2-dB taper with $h_s = 0.37$.

The background power received by the detector is proportional to the throughput. Compared to a $2Fl$ feedhorn-coupled detector, a $0.5Fl$ square pixel therefore receives less external power by a factor of 0.177/0.830 = 0.214; and a 1$Fl$ feedhorn-coupled detector receives less power by a factor of 0.370/0.830 = 0.446. The contributions of the external background photon noise to the overall *NEP*s are lower by the square roots of these numbers.

**3.5 Beam profiles on the sky**

Because the feedhorn is single-moded but the filled array is multi-moded, the beam profiles are calculated differently. For the feedhorn case, we use the standard radio-astronomical technique of computing the beam profile on the sky as the Fourier transform of the amplitude illumination profile at the telescope aperture. In the filled array case, it is more appropriate to regard the pixel as an absorbing area in the focal plane which couples to the intensity incident on it. The beam profile can then be computed as the two-dimensional convolution of the point spread function at the array focal plane with the pixel area.

To compare the beam widths for the filled and feedhorn cases, we assume for simplicity that the point spread function is diffraction limited and that the effects of primary obscuration can be neglected (to first order, departures from these assumptions will have equivalent effects on the beam profiles for the two cases). Figure 4 shows the FWHM beam-width (in units of $l/D$) as a function of edge taper for the feedhorn case and pixel size for the filled absorber case. For the feedhorn, increasing the edge taper broadens the beam on the sky because the outer portion of the aperture is less efficiently illuminated. For the filled pixel, the beam gets broader as the pixel gets larger because the profile is determined by the convolution of the PSF with the pixel. A feedhorn edge taper of 8 dB (2$Fl$), results in a beam only about 5% wider than the 0.5$Fl$ filled array beam-width, and for the 2-dB edge taper (1$Fl$), the beam is slightly narrower. We can therefore conclude that, in considering the angular resolution on the sky that can be achieved with the different array architectures, the differences in beam with are small for the $2F\lambda$ feedhorn, $1F\lambda$ filled, and $0.5F\lambda$ filled array cases. The choice of $1F\lambda$ filled array pixels does, however, result in a more significant broadening of the beam.

## 4. Signal-to-noise ratios and relative observing speed

### 4.1 Power levels on the detectors and overall *NEP*

The total radiant power absorbed by each pixel has four contributions as follows.

telescope background: $\qquad P_T = A W B_n(T_T) D n e_T \eta_s \eta_o \eta_d$ ; (3a)

atmospheric background: $\qquad P_A = A W B_n(T_A) D n e_A (1 - e_T) \eta_s \eta_o \eta_d$ ; (3b)

instrument cold box background: $\quad P_C = A W B_n(T_C) D n (1 - \eta_s) \eta_o \eta_d$ ; (3c)

the astronomical sky: $\qquad P_S = A W B_n(T_S) D n (1 - e_A)(1 - e_T) \eta_s \eta_o \eta_d$ . (3d)

We can write the total power as

$$P_{tot} = A W D n \eta_s \eta_o \eta_d \left[ B_{ext} + \frac{1 - \eta_s}{\eta_s} B_{int} \right], \qquad (4)$$

where

$$B_{ext} = e_T B_n(T_T) + e_A (1 - e_T) B_n(T_A) + (1 - e_A)(1 - e_T) B_n(T_S) \qquad (5)$$

is the external surface brightness (the sum of the telescope, atmosphere and sky contributions), and

$$B_{int} = B_n(T_C) \qquad (6)$$

is the internal surface brightness of the instrument cold box.

The photon noise limited *NEP*, referred to the power absorbed by the detector, is given by

$$NEP_{ph} = [2P_{tot}h\nu]^{1/2}. \tag{7}$$

The overall *NEP*, referred to the power absorbed by the detector, is

$$NEP_{tot} = [NEP_{ph}^2 + NEP_{det}^2]^{1/2}. \tag{8}$$

The performance of any system depends strongly on how the detector NEP compares to the photon noise limited NEP. We characterize this using the parameter γ, defined as $NEP_{det}/NEP_{ph}$. We then have

$$NEP_{tot} = NEP_{ph}[1+\gamma^2]^{1/2} = NEP_{det}\frac{[1+\gamma^2]^{1/2}}{\gamma}. \tag{9}$$

### 4.2 Measurement of sky intensity distribution (extended source)

For observation of a source extended with respect to the beam, the signal power, $P_S$, absorbed by a detector is that due to the astronomical sky brightness, as given by equation (3d). After an integration time, $t$, the signal-to-noise ratio, $s$, is

$$s = \frac{P_S(2t)^{1/2}}{NEP_{tot}} = \frac{AWB_\nu(T_S)D\nu(1-e_A)(1-e_T)\eta_s\eta_o\eta_d(2t)^{1/2}}{NEP_{tot}}. \tag{10}$$

For simplicity in comparing the different arrays, we can lump together into a constant, $K_1$, all of the quantities that are assumed to be independent of the detector type:

$$s = K_1\frac{\eta_s AWt^{1/2}}{NEP_{tot}}. \tag{11}$$

From (9) and (11), we can write the S/N achieved, relative to the $2F\mathbf{1}$ feedhorn array, with one of the other options in a given integration time for one detector as

$$\frac{\mathbf{s}}{\mathbf{s}_0} = \left[\frac{\mathbf{h}_s A W \mathbf{g}}{\mathbf{h}_{s0} A W_0 \mathbf{g}_0}\right]\left[\frac{1+\mathbf{g}_0^2}{1+\mathbf{g}^2}\right]^{1/2}. \tag{12}$$

The relative observing speed for one detector is proportional to $(\mathbf{s}/\mathbf{s}_0)^2$. If we consider maps of a given area made with multi-detector arrays, then for an array with $N/N_0$ times more detectors than the $2F\mathbf{1}$ array, the mapping speed will also be enhanced by a factor of $N/N_0$.

$$\frac{Speed}{Speed_0} = \frac{N}{N_0}\left[\frac{\mathbf{s}}{\mathbf{s}_0}\right]^2 = \frac{N}{N_0}\left[\frac{\mathbf{h}_s A W \mathbf{g}}{\mathbf{h}_{s0} A W_0 \mathbf{g}_0}\right]^2\left[\frac{1+\mathbf{g}_0^2}{1+\mathbf{g}^2}\right]. \tag{13}$$

In the completely detector-noise limited case ($\mathbf{g}, \mathbf{g}_0 \gg 1$), equation (13) shows that the speed ratio is just the ratio of the detector numbers multiplied by the square of the external throughput per detector (i.e., the square of the external photon rate). In the case where the noise is dominated by the external background ($\mathbf{g}, \mathbf{g}_0 \to 0$; $B_{int} = 0$), we have, from (4) and (7):

$$\left[\frac{\mathbf{g}}{\mathbf{g}_0}\right]^2 = \left[\frac{NEP_{ph0}}{NEP_{ph}}\right]^2 = \frac{P_{tot0}}{P_{tot}} = \frac{\mathbf{h}_{s0} A W_0}{\mathbf{h}_s A W}, \tag{14}$$

giving $$\frac{Speed}{Speed_0} = \frac{N}{N_0}\left[\frac{\mathbf{h}_s A W}{\mathbf{h}_{s0} A W_0}\right]. \tag{15}$$

Here the speed ratio is only linearly proportional to the throughput per detector because the dominant photon noise increases as the square root of the throughput.

### 4.3 Observation of a point source on-axis

When the flux density of a point source is being measured, the coupling efficiency of the individual detectors to the point spread function is important. The background power and photon noise limited *NEP* are as given above in equations (4) and (7). The signal power absorbed by the on-axis detector is.

$$P_S = S_n A_T (1 - e_A)(1 - e_T) D n h_A h_o h_d . \tag{16}$$

Equation (10) then gives

$$s = \frac{S_n A_T (1 - e_A)(1 - e_T) D n h_A h_o h_d (2t)^{1/2}}{NEP_{tot}} . \tag{17}$$

We can again combine together all of the quantities that are assumed to be independent of the array architecture:

$$s = K_2 \frac{h_A t^{1/2}}{NEP_{tot}} , \tag{18}$$

where $K_2$ is another constant.

For the feedhorn arrays and the $1F\lambda$ filled array, the main beams of the nearest neighbor detectors are too far off-axis to collect any appreciable power from a point source, so the measurement is by the on-axis detector only. No jiggle pattern is needed, and the total available integration time is used on-source. In the case of the $0.5F\lambda$ filled array, the signals from the eight neighboring pixels can be co-added to improve the S/N, with the signals weighted in accordance with the beam pattern. We calculate the available improvement in S/N as follows. All nine pixels are assumed to have the same noise level. The relative signal in each of the pixels is derived by computing the volume under the portion of the diffraction-limited PSF intercepted by that pixel. Each of the nine pixel measurements is taken as an independent estimate of the source signal. Normalizing the S/N in the central pixel to unity, the overall S/N is improved by a factor given by the quadrature sum of the relative S/N values in all nine pixels. The result of this calculation is a S/N improvement factor of 1.58 compared to value given by equation (17) for the on-axis detector alone. The relative observing speed ratio is given by the square of the S/N ratio. Comparing the observing speed achieved for a given integration time (for the on-axis detector alone) with that for the $2Fl$ case, we therefore have from (9) and (18)

$$\frac{Speed}{Speed_0} = f\left[\frac{s}{s_0}\right]^2 = f\left[\frac{h_A g}{h_{A0} g_0}\right]^2 \left[\frac{1+g_0^2}{1+g^2}\right]. \quad (19)$$

where the factor $f = 1.58^2 = 2.5$ for the $0.5Fl$ case, and $f = 1$ for the other cases.

**4.4 Seven-point observation of a point source with the feedhorn option**

Should the pointing accuracy or knowledge of the source position be such that one cannot rely on blind pointing to make an accurate measurement of a point source, then, for the feedhorn option, it is necessary to carry out a small map around the source. This reduces the S/N for a given integration time because the time must be shared between several positions. The most efficient approach is to perform a five, seven, or nine-point map in which the nominal position and a set of adjacent positions around it are visited in turn by a detector. The spacing, $\Delta q$, should be larger than the maximum expected pointing error or the uncertainty in the source position, but smaller than the Nyquist sampling interval. If such is not possible, then a fully sampled jiggle map must be carried out as for an extended source. Here we assume that a 7-point observation is carried out, with the signals from the seven map points co-added to derive the total flux density, and compare the final S/N with the S/N that would be obtained if the pointing were good enough to devote all of the integration time to one position.

Compared to the value for the central position alone, the total signal is increased by a factor of

$$1 + 6\exp\left[-\left[\frac{2[\ln(2)]^{1/2} Dq}{q_{FWHM}}\right]^2\right].$$

The noise per position is increased by a factor of $7^{1/2}$ compared to that for a single long integration because the integration time is shared between the seven positions; and the final noise level is increased by a further $7^{1/2}$ through the co-addition of the seven signals. The final S/N is therefore reduced by a factor of

$$7\left[1 + 6\exp\left[-\left[\frac{2[\ln(2)]^{1/2} Dq}{q_{FWHM}}\right]^2\right]\right].$$

This square of this factor, representing the loss in observing speed, is plotted against the step size (normalized to the beam FWHM) in Figure 5. For a typical value of $\Delta q = 0.3$, performing the 7-point results in an observing speed reduction by a factor of 0.66.

**4.5 Extraction of point sources from maps**

An important scientific goal for imaging submillimeter instruments is to carry out surveys of unexplored areas of sky and to detect faint point sources in the maps. Such observations can be made either by taking data on-the-fly while scanning the telescope continuously, or by raster scanning. In the former case, spatial modulation of the signal can be implemented either by chopping the field of view or by choosing a scan rate such that the signal frequencies are above the $1/f$ knee of the system. In order to extract optimally all the necessary information on the sky and on the characteristics of the detector system, a "dithering" scheme involving more complex modulation of the beam may be implemented [19].

If modulation is implemented by scanning alone (without chopping or dithering), then the scan rate must also be such that the beam crossing time is long compared to the detector time constant to avoid loss of angular resolution in the scan direction. Chopping and/or dithering can also be carried out in making maps. In these modes, the individual detector signals must be explicitly recorded: differencing prior to recording will add significantly to the confusion noise due to the larger effective beam area, increasing the confusion limit by a factor of ~ 1.6.[20]

Scanning observations with feedhorn arrays produce a fully sampled image of the sky covered by each individual scan provided the scan direction is chosen to give the necessary overlap between the beams. For chopped observations with feedhorn arrays, a "jiggle-pattern" must be performed to get a fully sampled map. Notwithstanding any differences in the observing modes, it is possible to determine generically the relative mapping speeds of the feedhorn and filled array architectures. Consider a fully-sampled map of a given area with a certain total integration time and a fixed spacing between the samples. In practice the sampling grid may be more complicated, but that will have little or no effect on the result of this comparison. Regardless of the exact details of the observing modes or the size of the map area in relation to the array field of view, in the filled array case we have $N/N_0$ times more integration time per sample than for the feedhorn case. If the map is critically sampled ($0.5\lambda/D$ spacing) then the filled array generates it purely by spatial multiplexing: the array fully samples the sky such that all necessary spatial samples are taken at the same time. The feedhorn array generates the final image by a combination of spatial and temporal multiplexing: the array samples multiple points on the sky, but does not produce a fully sampled image with a single telescope pointing - sequential observations are also needed to complete the image. If the map is over-sampled (say, $0.25 l/D$) then both arrays use a combination of spatial and temporal multiplexing. In any case, what determines the mapping speed is the final S/N per map point, which is considered below. For simplicity, we assume that the map contains a point source that happens to be centered on one of the map points, and that the small difference in the beam widths between the filled array and feedhorn-coupled systems can be neglected.

Consider a map with integration time $t$ per map position. Let the map contain an isolated point source coincident with one of the map points, with signal $S_a$ at that position. Let $\Delta S$ be the noise level in each map position. Let $s_a = S_a/\Delta S$ be the S/N for the central position (given by equation 17), and let $n$ appropriate map points be co-added to enhance S/N on the point source.

The signal in pixel $i$ is

$$S_i = S_a \exp\left[-\left[\frac{2[\ln(2)]^{1/2} Dq_i}{q_{FWHM}}\right]^2\right], \tag{20}$$

where $\Delta q_i$ is the angular offset from the source. The final signal level is

$$S_{tot} = \sum S_i = K_3 S_a, \tag{21}$$

where $K_3$ is a constant that depends on the grid spacing and $q_{FWHM}$. The final noise level is

$$\Delta S_{tot} = n^{1/2} \Delta S, \tag{22}$$

and the final S/N is thus

$$\boldsymbol{s} \;=\; \frac{K_3}{n^{1/2}}\boldsymbol{s}_{\text{a}} \;=\; \frac{K_2 K_3 \boldsymbol{h}_{\text{A}} t^{1/2}}{n^{1/2} NEP_{\text{tot}}}. \tag{23}$$

For an array with $N/N_0$ times more detectors than the $2F\boldsymbol{l}$ case, the integration time per point is $N/N_0$ times longer. Taking the square of the ratio of the S/N values gives the relative mapping speed with respect to the $2F\boldsymbol{l}$ feedhorn case:

$$\frac{Speed}{Speed_0} \;=\; \frac{N}{N_0}\left[\frac{\boldsymbol{s}}{\boldsymbol{s}_0}\right]^2 \;=\; \frac{N}{N_0}\left[\frac{\boldsymbol{h}_{\text{A}}\boldsymbol{g}}{\boldsymbol{h}_{\text{A}0}\boldsymbol{g}_0}\right]^2 \left[\frac{1+\boldsymbol{g}_0^{\,2}}{1+\boldsymbol{g}^2}\right] \tag{24}$$

which is the same as equation (19) for a point source on-axis except that the factor $f$ is replaced by $N/N_0$.

## 5. Observing speed comparison

### 5.1 Case of zero instrument background

In this section we consider the case in which the internal background due to the instrument cold box can be neglected (i.e., $B_{\text{int}} \approx 0$). This is a reasonable assumption for all of the options at far infrared and submillimeter wavelengths but is not necessarily valid for filled array detectors at millimeter wavelengths. The case of non-negligible instrument background is considered below in Section 5.2. The array parameters derived in Section 3 for the four cases are listed in Table 1. These have been used to calculate the relative observing speed using the equations derived in Section 4.

(a) Purely background-limited case

For completely background limited detectors, the parameter $\boldsymbol{g}$ is zero (i.e., $NEP_{\text{det}} \ll NEP_{\text{ph}}$) in all cases. The observing speed ratios, as defined in Section 4, are then as given in Table 2. For extended source observation, the filled arrays are the fastest options, with no difference in principle (barring the small difference in angular resolution) between the $0.5F\lambda$ and $1F\lambda$ cases. The potential enhancement in observing speed over the $2F\lambda$ feedhorn case is about a factor of 3. For point source extraction from a map, the $0.5F\lambda$ filled array is faster than any of the other options, with a speed advantage of 3.5 compared to the $2F\lambda$ feedhorn case. The two $1F\lambda$ options have comparable speed at around a factor of 2 faster than the $2F\lambda$ case. For observation of a known point source with good pointing accuracy, the $2F\lambda$ feedhorn option is significantly better than any of the others. If a seven point needs to be implemented, then the sensitivity is still comparable to that of the $0.5F\lambda$ filled array case. The $1F\lambda$ feedhorn case is poorly optimized for this kind of observation.

(b) Partly detector noise limited case

A more realistic assumption is that the detector *NEP* is finite, so that $g$ is non-zero. The observing speed ratios are plotted vs. $g$ in Figures 6, 7 and 8, covering the range $\gamma = 0 - 1$, where 0 corresponds to the purely background-limited case and 1 corresponds to $NEP_{det} = NEP_{ph.}$

For an extended source (Figure 6) the filled array speed advantage holds up best for the $1F\lambda$ case because the background level is similar to that for the $2F\lambda$ feedhorns. The speed advantage for the $0.5F\lambda$ array declines to around a factor of 2 for $\gamma = 1$. This is because the lower background on the filled array detectors means that, as $NEP_{det}$ increases, the departure from background limited operation is greater than for the feedhorn-coupled detectors: to take advantage of the potential observing speed advantage, more sensitive detectors are needed. For point source extraction from a map (Figure 7), the advantage of the $0.5F\lambda$ option again decreases to around a factor of 2 for $g = 1$. The $1F1$ options have comparable sensitivity over the range of backgrounds. For observation of a point source on-axis (Figure 8), the $2F1$ option is the fastest, even if a 7-point has to be executed, and the relative advantage becomes slightly greater as $NEP_{det}$ increases. As for the other observations, the relative performance of the $1F1$ filled array remains fairly uniform over the range of backgrounds.

## 5.2 Non-zero instrument background

Because filled array pixels have a much larger throughput than feedhorn coupled detectors, they are prone to collect much more background power from the instrument itself. At millimeter wavelengths, this can become a significant or even dominant contribution to the photon noise unless the cold box surrounding the detectors is maintained at a very low temperature. Important potential applications are cosmic background radiation observations and deep extragalactic surveys in the 2 - 3 mm region. To examine the impact of the instrument background, we consider the dependence of the observing speed on $B_{int}/B_{ext}$, the ratio of the instrument cold box brightness to the external brightness.

The sensitivity of the observing speed ratios (equations 13, 19, 24) to $B_{int}/B_{ext}$ can be examined by noting that, from the definition of $\gamma$ and equations (4) and (7) we can write

$$\left[\frac{g}{g_0}\right]^2 = \frac{P_{tot0}}{P_{tot}} = \left[\frac{AW_0}{AW}\right]^2 \left[\frac{h_{s0} + (1+h_{s0})\left(\frac{B_{int}}{B_{ext}}\right)}{h_s + (1+h_s)\left(\frac{B_{int}}{B_{ext}}\right)}\right]. \tag{25}$$

Here we consider the case of an extended source in the general background limited case ($\gamma, \gamma_0 \to 0$; $B_{int} \neq 0$), where equation (13) becomes

$$\frac{Speed}{Speed_0} = \frac{N}{N_0}\left[\frac{h_s}{h_{s0}}\right]^2\left[\frac{AW}{AW_0}\right]\left[\frac{h_{s0} + (1+h_{s0})\left(\frac{B_{int}}{B_{ext}}\right)}{h_s + (1+h_s)\left(\frac{B_{int}}{B_{ext}}\right)}\right]. \quad (26)$$

This speed ratio is plotted vs. $B_{int}/B_{ext}$ in Figure 9, with all curves normalized to the case of the 2*F1* array with $B_{int}/B_{ext} = 0$. The 2*F1* case is very insensitive to the instrument cold box emission because the highly directional angular response of the feedhorn illuminates the pupil preferentially. Even the broader 1*F1* feedhorn beam results in only a small degradation in speed provided the internal:external brightness ratio is less than around 10%. However, for the filled arrays, the speed advantage is rapidly eroded by instrument background photon noise for $B_{int}/B_{ext} > 0.1\%$, and has vanished entirely for $B_{int}/B_{ext} > 2\%$. The mapping speed advantage for the other kinds of observation is similarly sensitive to the background instrument background (but the curves for the two filled array options are identical only for the particular case shown here of an extended source with zero detector *NEP*).

Figure 10 shows the brightness ratio as a function of cold box temperature for two typical examples:

(i) a ground-based experiment at $\lambda = 3$ mm with $T_T = 280$ K, $e_T = 0.05$, $T_A = 250$ K, $e_A = 0.05$, $T_S = 2.73$ K;

(ii) a space-borne experiment at $\lambda = 2$ mm with $T_T = 60$ K, $e_T = 0.01$, $T_A = 0$, $T_S = 2.73$ K.

Retaining a factor of 2 advantage in speed for the filled array options requires $B_{int}/B_{ext} < 0.005$, so the temperature of the detector enclosure must be maintained at less than ~ 1.3 K (ground-based) or ~ 1 K (space experiment). We note that these results are optimistic for the filled arrays in that they assume that the detector NEP is zero and that stray light is negligible. In addition, the strong temperature dependence of the cold-box background in the Wien region of the black-body spectrum would dictate either lowering the temperature further or implementing very precise temperature control.

## 6. Conclusions

In addition to operational advantages, filled bolometer arrays offer, for a given field area, improved sensitivity for mapping observations at the expense of larger detector numbers. In order to take complete advantage of the fully-sampled filled array architecture, the detectors must be operated close to the background limit. In that limiting case, the fully-sampled filled array is 3.5 times faster than the traditional $2F\lambda$ feedhorn array for the extraction of point sources from mapping observations. For a given number of detectors, feedhorn-coupled architectures provide better mapping speed than filled arrays, at the expense of a larger field of view. The $2F\lambda$ feedhorn provides the best possible sensitivity for observations of a known point source. Given the complexity of bolometer instruments, practical considerations such as stray light and RF suppression, multiplexing capabilities, power dissipation, available data rate, instrument cryogenic design and temperature stability, etc., may be important in deciding between the various options.

**Acknowledgements**

We are grateful for helpful discussions with Martin Caldwell, Jason Glenn, Harvey Moseley, Anthony Murphy, Seb Oliver, and Bruce Swinyard. We would also like to thank the referees for their comments which allowed us to improve the paper significantly.

Table 1. Summary of array parameters for the four architectures (with zero instrument background)

|  | Filled array | | Feedhorn array | |
| --- | --- | --- | --- | --- |
| **Pixel size** | **0.5F$\lambda$** | **1.0F$\lambda$** | **1.0F$\lambda$** | **2.0F$\lambda$** |
| $N/N_0$ | 13.9 | 3.46 | 4 | 1 |
| $\eta_A$ | 0.162 | 0.495 | 0.350 | 0.700 |
| $A\Omega$ (units of $\lambda^2$) | $0.226F^2\pi$ | $0.903F^2\pi$ | 1 | 1 |
| Edge taper (dB) | 0 | 0 | 2 | 8 |
| $\eta_s$ | $1/(4F^2)$ | $1/(4F^2)$ | 0.370 | 0.830 |
| Beam FWHM (units of $\lambda$/D) | 1.07 | 1.22 | 1.05 | 1.13 |
| Background power per pixel | 0.214 | 0.855 | 0.446 | 1 |
| Normalized $NEP_{ph}$ | 0.462 | 0.925 | 0.668 | 1 |
| $g/g_0$ | 2.16 | 1.08 | 1.50 | 1 |

Table 2. Relative observing speed (purely background limited; zero instrument background)

|  | Filled array | | Feedhorn array | |
| --- | --- | --- | --- | --- |
|  | **0.5F$\lambda$** | **1.0F$\lambda$** | **1.0F$\lambda$** | **2.0F$\lambda$** |
| Extended source | 2.97 | 2.97 | 1.78 | 1 |
| Point source extraction from map | 3.48 | 2.03 | 2.24 | 1 |
| Point source photometry (on-axis) | 0.625 | 0.587 | 0.561 | 1 |
| Point source photometry (7-point) | - | - | 0.369 | 0.658 |

**Figure captions**

Figure 1: Jiggle pattern needed to achieve a fully sampled map with square-packed $2F\lambda$ feedhorns (in the case of hexagonal close packing the jiggle pattern is slightly different but 16 steps are still needed).

Figure 2: Filled array and feedhorn aperture efficiencies vs. pixel side (square filled array pixel) or horn aperture diameter (smooth-walled conical horn).

Figure 3: Feedhorn spillover efficiency $h_s$ vs. Gaussian edge taper of the telescope pupil.

Figure 4: FWHM beam-width, in units of $\lambda/D$, vs. edge taper, in dB, for a feedhorn coupled detector (upper plot), and pixel size in units of $F\lambda$ for a filled absorber (lower plot).

Figure 5: Observing speed loss vs. 7-point angular step (normalized to beam FWHM).

Figure 6: Mapping speed vs. γ for an extended source observation, normalized to the $2F\lambda$ feedhorn case (zero instrument background).

Figure 7: Observing speed vs. γ for point source extraction from a map, normalized to the $2F\lambda$ feedhorn case (zero instrument background).

Figure 8: Observing speed for a point source on-axis, normalized to the 2F$\lambda$ feedhorn case (zero instrument background)

Figure 9: Observing speed vs. the ratio of the internal:external brightnesses for purely background-limited observation of an extended source. All curves are normalized to the 2F$\lambda$ feedhorn case with $B_{int}/B_{ext} = 0$.

Figure 10: Internal:external brightness ratio vs. cold box temperature for the ground-based and space-borne experiments discussed in the text.

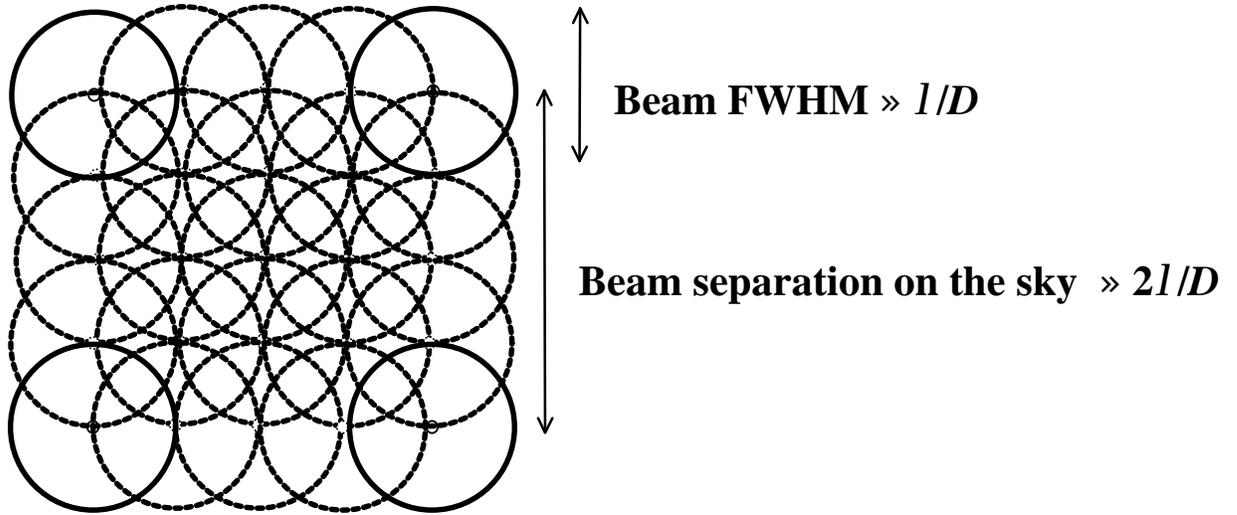

Figure 1

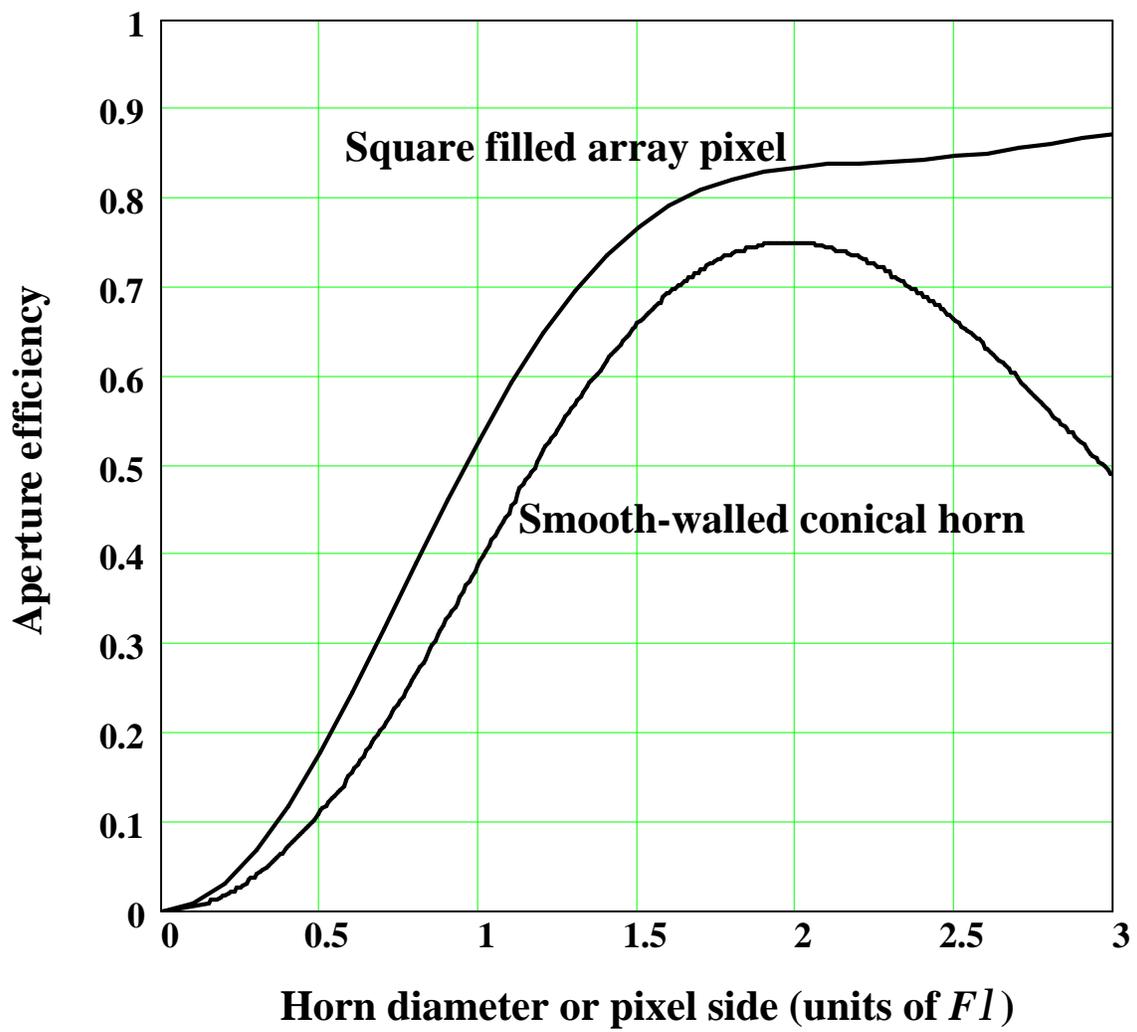

Figure 2

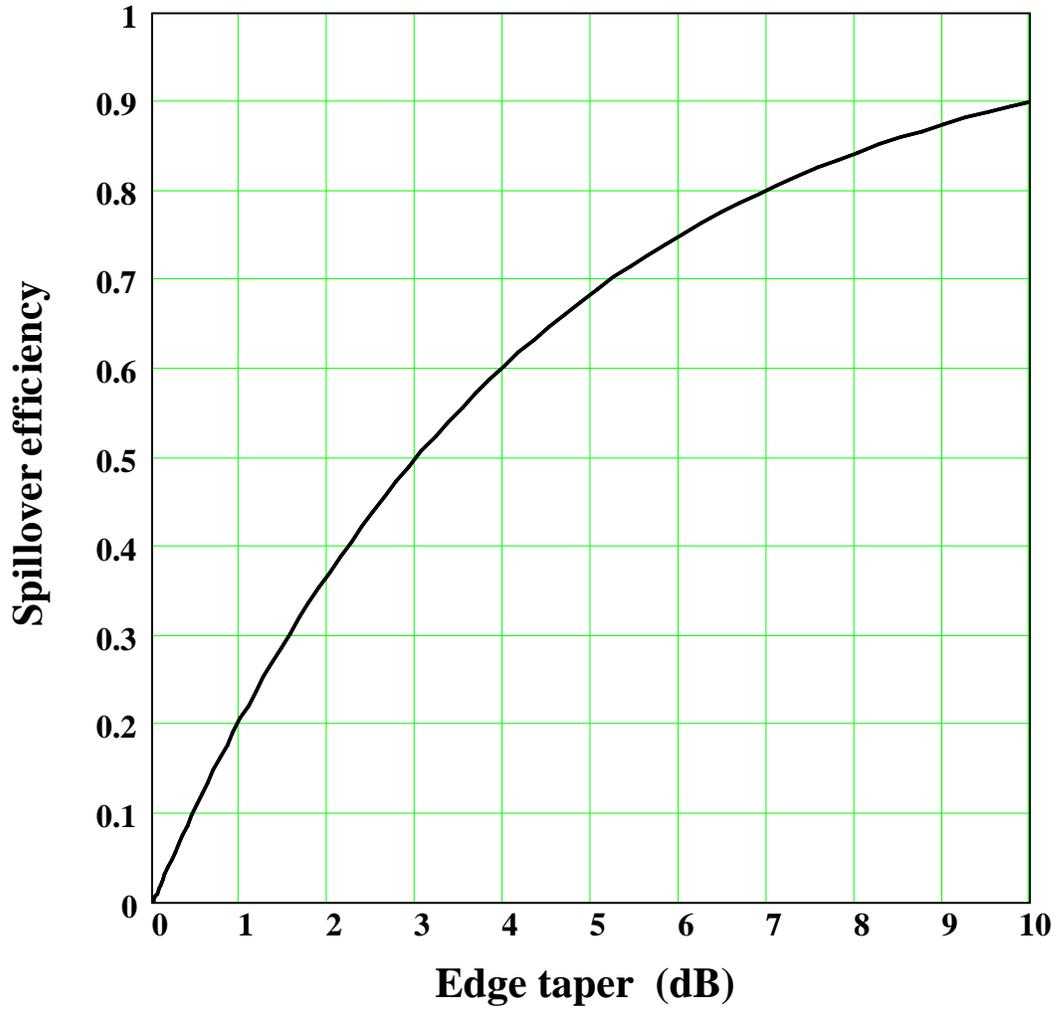

Figure 3

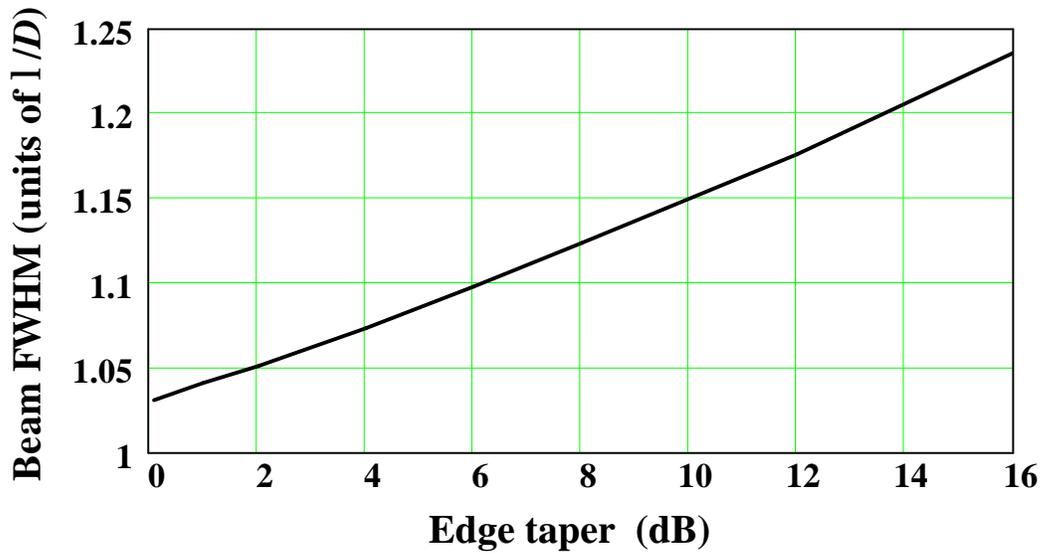
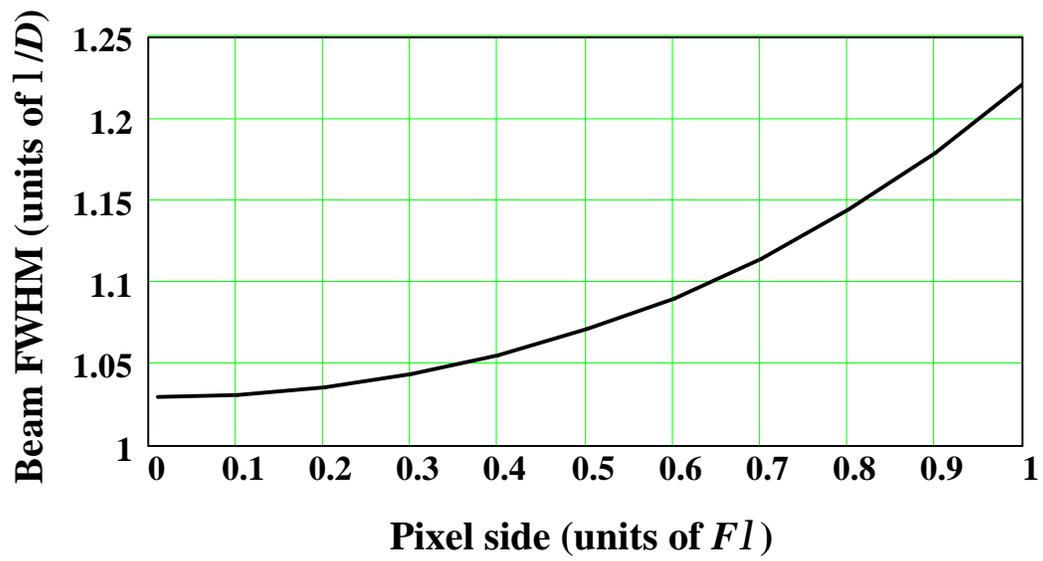

Figure 4

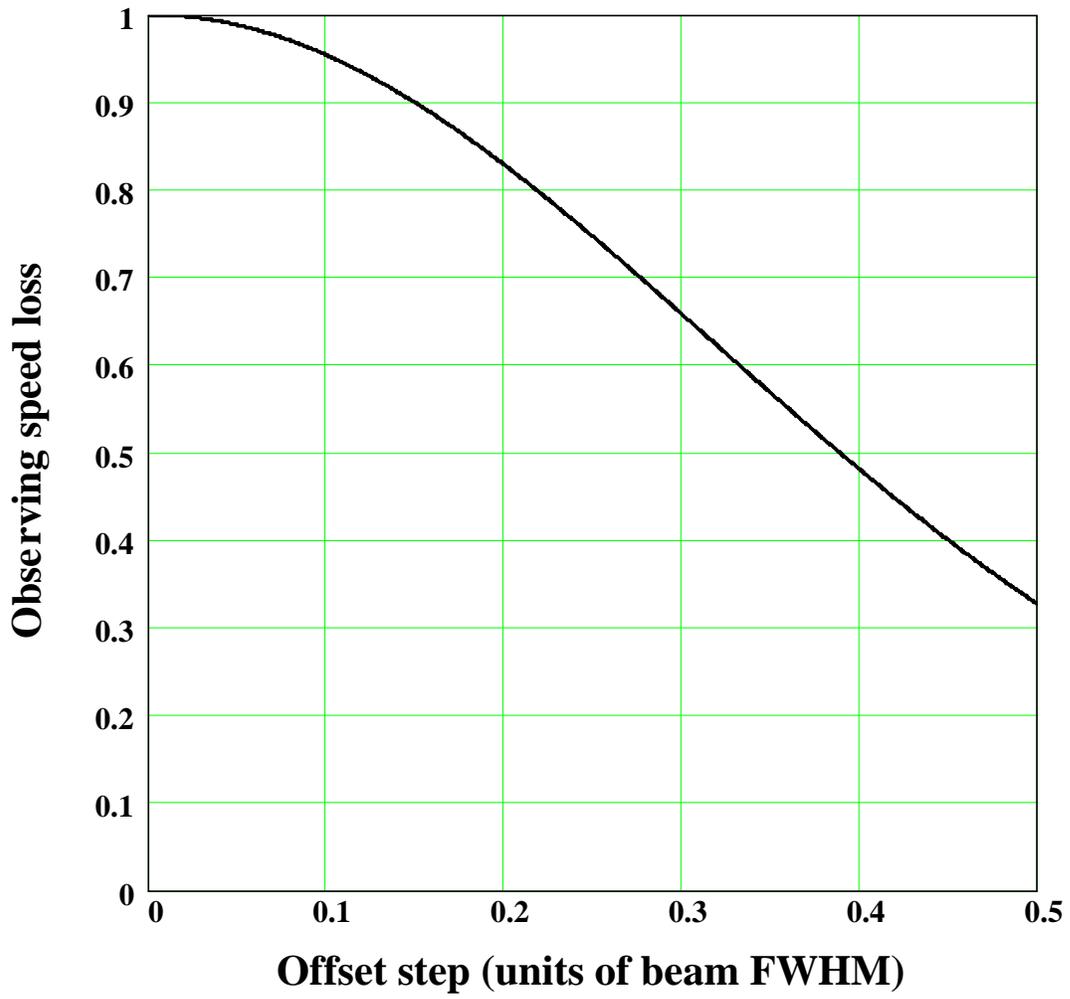

Figure 5

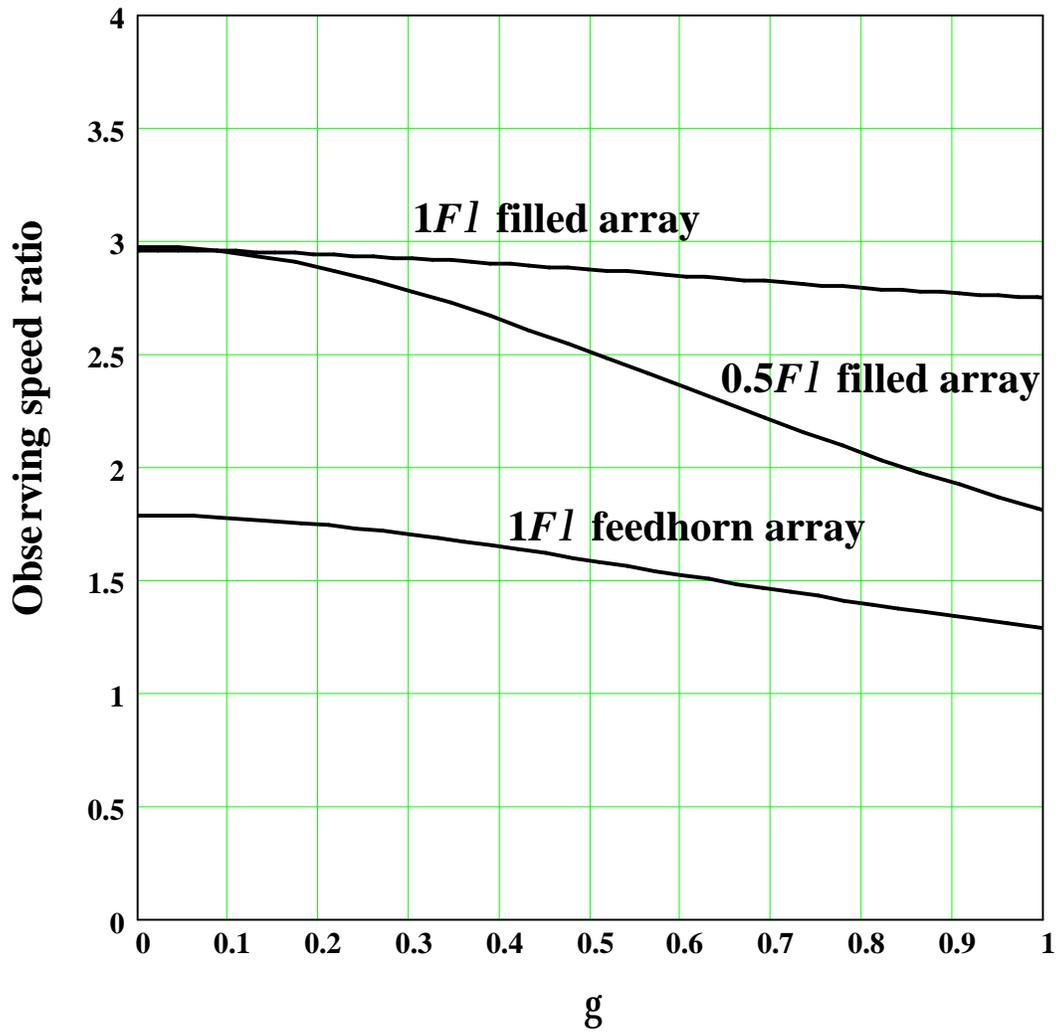

Figure 6

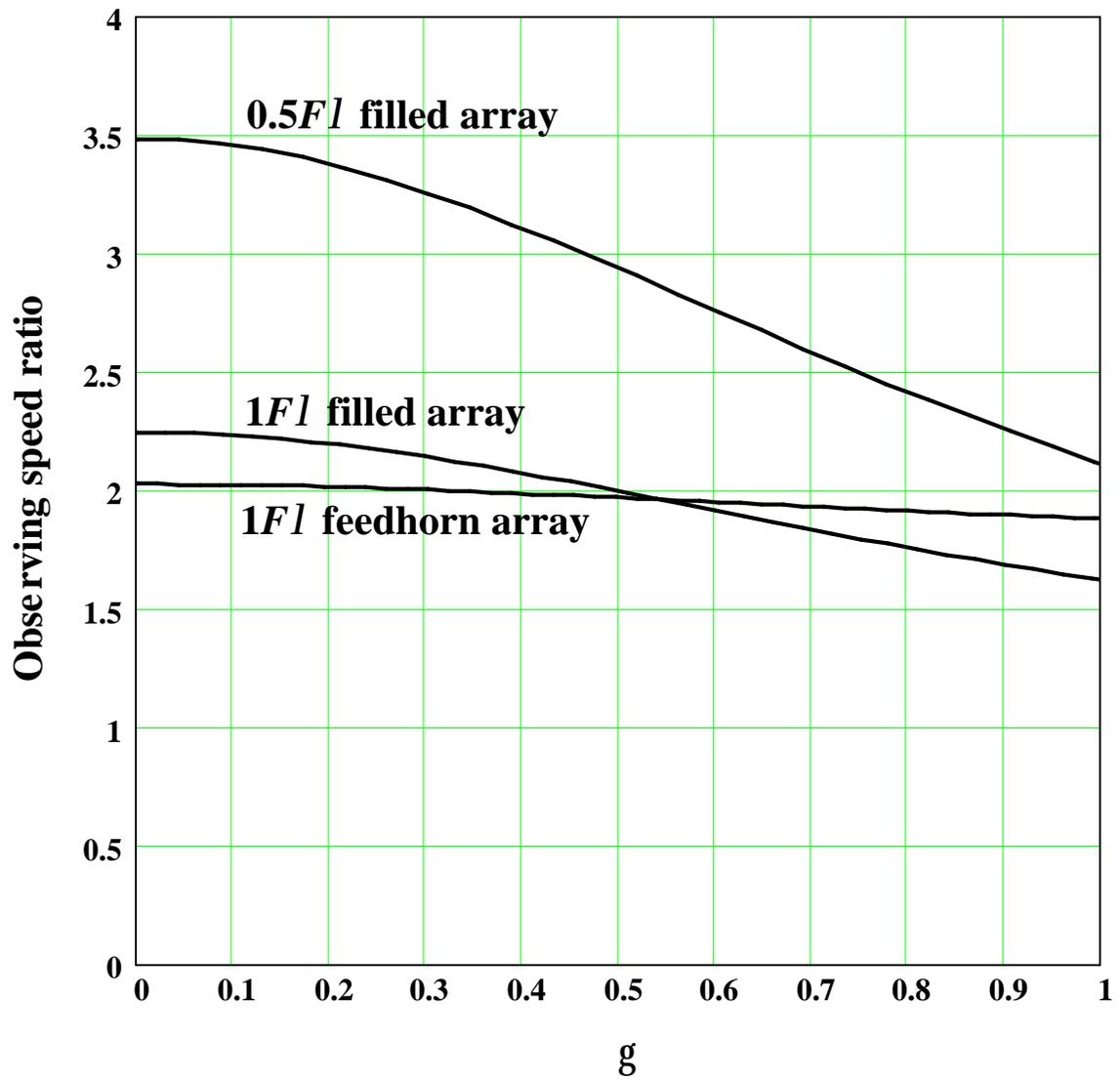

Figure 7

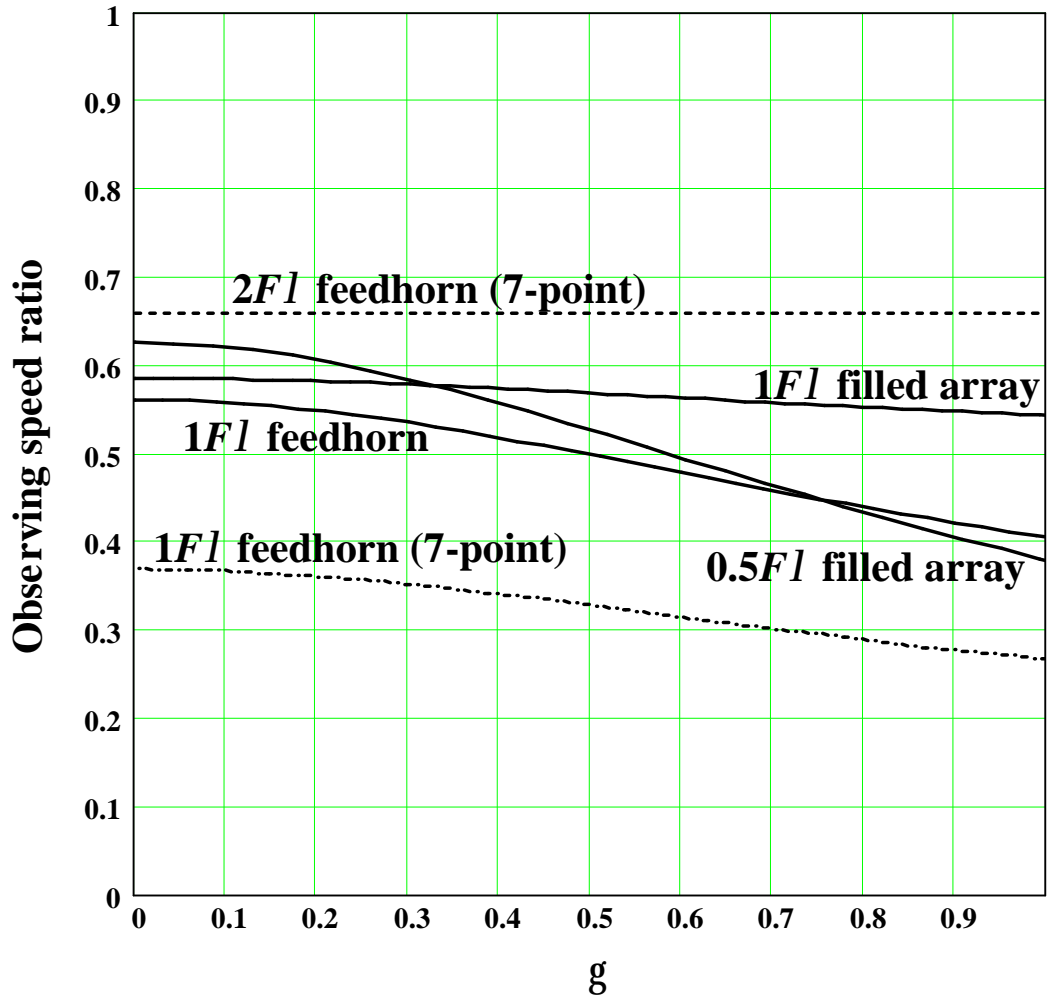

Figure 8

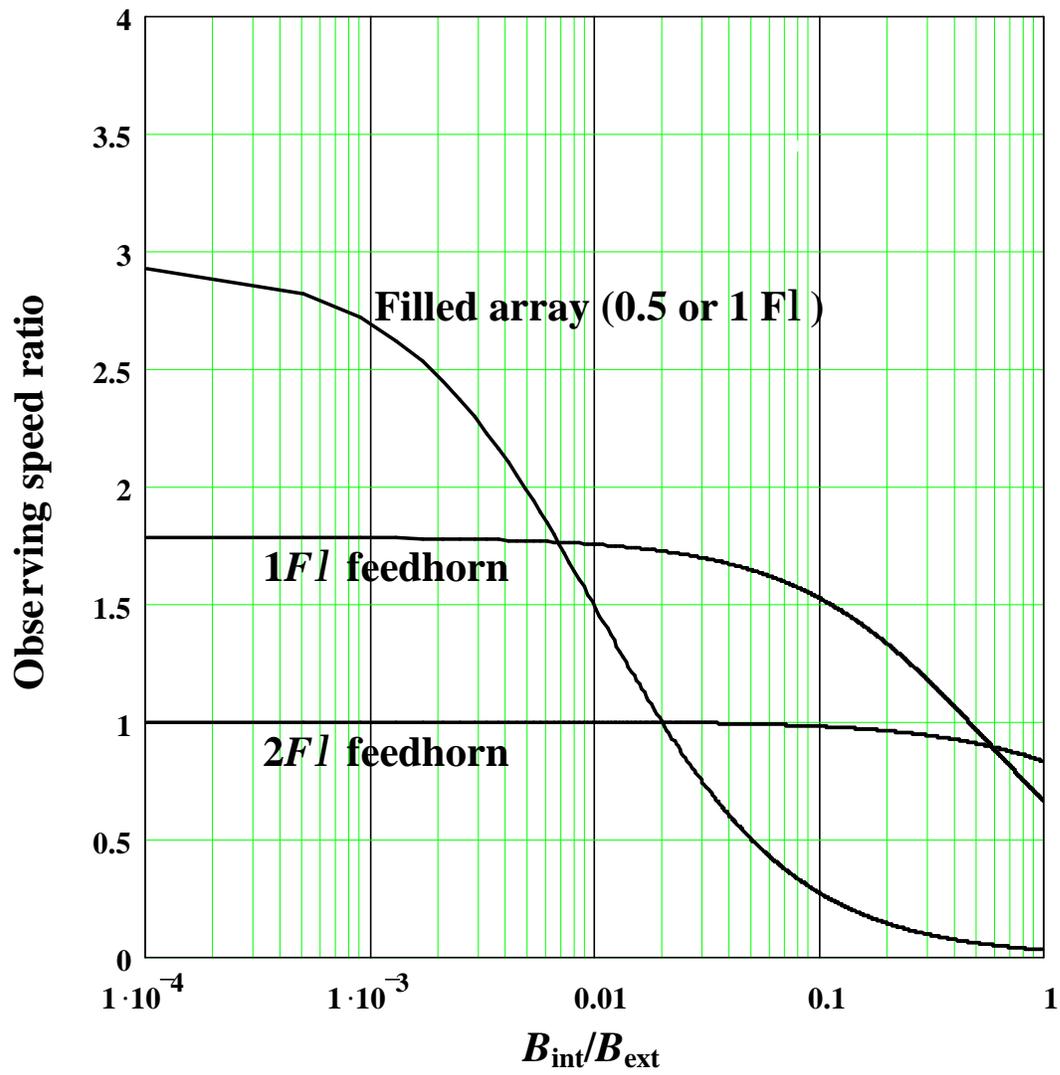

Figure 9

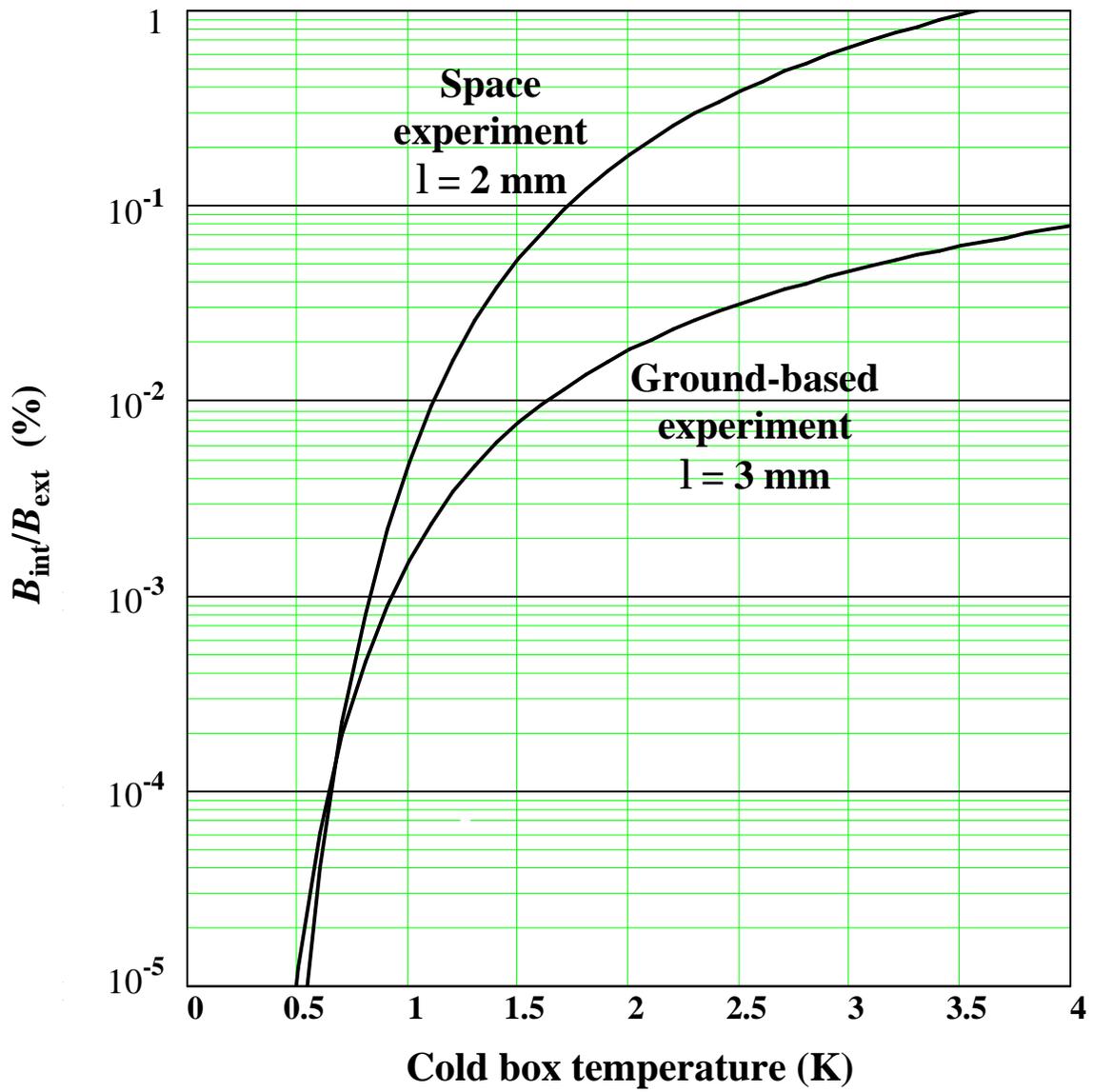

Figure 10